# Enigmatic Aspects of the Early Universe: Possibility of a 'Pre-Big Bang Phase'!

C Sivaram and Kenath Arun

Indian Institute of Astrophysics, Bangalore

**Abstract:** In this paper it is suggested that inclusion of mutual gravitational interactions among the particles in the early dense universe can lead to a 'pre-big bang' scenario, with particle masses greater than the Planck mass implying an accelerating phase of the universe, which then goes into the radiation phase when the masses fall below the Planck mass. The existence of towers of states of such massive particles (i.e. multiples of Planck mass) as implied in various unified theories, provides rapid acceleration in the early universe, similar to the usual inflation scenario, but here the expansion rate goes over 'smoothly' to the radiation dominated universe when temperature becomes lower than the Planck temperature.

It is now generally accepted that the universe went through a very hot dense phase and many of the predictions of this standard evolutionary model, including the anisotropies [1, 2] and the light element abundances [3] are now well established [4]. However there are still some enigmatic [5] aspects of the earliest phase of the universe, which are only partially explained by models such as those involving inflation. [6]

In the standard evolutionary big bang cosmological model, the radius of the expanding universe is related to the corresponding temperature of the radiation by:

$RT = constant$ ... (1)

For the present epoch the temperature is of the order of $3^0$ Kelvin and the scale factor (Hubble radius) of the universe is given as:

$R = \frac{c}{H_0} \sim 10^{28} cm$ ... (2)

This implies $RT = 3 \times 10^{28} cmK$, which remains constant.

At the era when matter and radiation were of comparable energy densities, which also happens to be around the recombination era ($T \approx 4000K$)

$R \approx 10^{25} cm$ ... (3)

The expansion rate of the universe is given by:

$$\frac{\dot{R}^2}{R^2} = \frac{8\pi G\rho}{3} \qquad \ldots (4)$$

where the density is given as:

$$\rho = \frac{m}{\frac{4}{3}\pi\left(\frac{h}{mc}\right)^3} = \frac{m^4 c^3}{\frac{4\pi}{3} h^3} \qquad \ldots (5)$$

(since in the early epoch the particles are separated by the Compton length)

The expansion rate is then given by:

$$H^2 = \frac{\dot{R}^2}{R^2} = \frac{2Gm^4 c^3}{h^3} = \frac{2m^2 c^4}{h^2}\frac{m^2}{m_{Pl}^2} \qquad \ldots (6)$$

(where $m_{Pl} = \left(\frac{hc}{G}\right)^{1/2}$ is the Planck mass)

However, if we also include the mutual gravity between the particles, an additional term in the energy density, i.e. $\rho_G$, gets introduced, which is given by: [7]

$$\rho_G = \frac{Gm^2}{r^4}\bigg/c^2 \qquad \ldots (7a)$$

Where: $r = \frac{h}{mc}$

Thus: $\rho_G = \frac{Gm^6 c^2}{h^4} \qquad \ldots (7b)$

This is the binding energy density. So the expansion rate can be written as:

$$\frac{\dot{R}^2}{R^2} = \frac{8\pi G m^4 c^3}{3h^3}\left(1 - \frac{Gm^2}{hc}\right) \qquad \ldots (8)$$

Similarly the acceleration equation is given by:

$$\frac{\ddot{R}}{R} = -4\pi G(\rho + \rho_G) = -\frac{4\pi G m^4 c^3}{3h^3}\left(1 - \frac{Gm^2}{hc}\right) = -\frac{4\pi G m^4 c^3}{3h^3}\left(1 - \frac{m^2}{m_{Pl}^2}\right) \qquad \ldots (9)$$

At $m = m_{Pl}$, the two terms become comparable and we have: $\ddot{R} = 0$

At $m > m_{Pl}$, the acceleration becomes positive, i.e. the motion can be reversed (starts expanding), in other words initiating an expanding phase.

It is also possible, as suggested in string theory as well as in quantum gravity and Kaluza-Klein (KK) theories, that mass spectrum is in multiples of Planck mass, i.e., $m >> n\left(\frac{hc}{G}\right)^{1/2}$.

Therefore the above criteria (i.e. $m > m_{Pl}$) can be satisfied, i.e. we can have a spectrum of states starting from the Planck mass (so called Tower of Spin-2 States). The Planck mass, as such, it is to be remembered comes from quantum considerations and is not classical. It is only the ground state. One could have multiples of h to get excited states like in the case of oscillators.

So this is a possible scenario for reversing the collapse or initiating an expansion phase. The acceleration equation in terms of temperature is given by:

$$\frac{\ddot{R}}{R} = -\frac{4\pi G m^4 c^3}{3h^3}\left(1 - \frac{T^2}{T_{Pl}^2}\right) \quad \ldots (10)$$

Where $T_{Pl}$ corresponds to $m_{Pl} c^2$.

At temperatures, $T > T_{Pl}$ in the early universe, the expansion is repulsive giving the inflation scenario, till the temperature become equal to the Planck temperature. As the temperature lowers further, the expansion rate decreases and goes into the radiation era for $T < T_{Pl}$.

So we effectively have a 'pre-big bang' scenario, where particles' masses $>> m_{Pl}$, which from equation (9), implies an accelerating phase of the universe (i.e. $\ddot{R}$ is positive), which goes into the radiation phase, when (from equation (10)) $m << m_{Pl}$. There is a lot of justification for the existence of a tower of states ($> m_{Pl}$), from Klein-Kaluza and superstring theories. The existence of such states also provides rapid acceleration in the early universe (similar to the usual inflation scenario), but here the expansion rate goes over 'smoothly' to the radiation dominated universe when $T << T_{Pl}$.

**Reference:**
1. John C Mather, Rev. Mod. Phys., 79, 1331, 2007
2. G F Smoot, Rev. Mod. Phys., 79, 1349, 2007